\documentclass{nature}
\usepackage{graphicx}
\makeatletter
\let\saved@includegraphics\includegraphics
\AtBeginDocument{\let\includegraphics\saved@includegraphics}

\makeatother
\usepackage{mathtools}
\usepackage{amsmath,amssymb}
\usepackage{amsfonts}
\usepackage{braket}
\usepackage{soul}
\usepackage{color}
\definecolor{orange}{rgb}{1,0.5,0}

\mathchardef\Re="023C
\mathchardef\Im="023D

\title{Long living carriers in a strong electron-phonon interacting two-dimensional doped semiconductor}

\author{Peio Garcia-Goiricelaya$^{1,2}$, Jon Lafuente-Bartolome$^{1,2}$, Idoia G. Gurtubay$^{1,2}$ \& Asier Eiguren$^{1,2}$}

\begin{document}

\maketitle

\begin{affiliations}
\item Materia Kondentsatuaren Fisika Saila, University of the Basque Country UPV/EHU, 48080 Bilbao, Basque Country, Spain.
\item Donostia International Physics Center (DIPC), Paseo Manuel de Lardizabal 4, 20018 Donostia-San Sebasti\'{a}n, Spain.
\end{affiliations}

\begin{abstract}
Carrier doping by the electric field effect has emerged recently as an ideal route 
for monitoring many-body physics in two-dimensional (2D) materials where
the Fermi level is tuned in a way that -indirectly- the strength of the interactions 
can also be scanned \cite{controlmb1,controlmb2}.
The possibility of systematic doping in combination 
with high resolution photoemission has allowed to uncover a genuinely many-body 
electron spectrum in single-layer MoS$\mathbf{_{2}}$ transition metal dichalcogenide, resolving three 
clear quasi-particle states,  where only one state should be expected
if the electron-phonon interaction vanished \cite{Kang2018}. 
Our analysis combines first-principles and consistent complex plane analytic approaches 
and brings into light the presence and the physical origin of two gaps and the three 
quasi-particle bands which are unambiguously
 present in the photoemission spectrum. One of these states, 
though being strongly interacting with the accompanying virtual  phonon cloud, 
presents a notably long lifetime which is  
an appealing property when trying to understand and take advantage of many-body interactions to modulate 
the transport properties~\cite{scmos21,scmos22,scmos23,scmos24}.

\end{abstract}

\begin{figure*}[ht!]
\begin{center}
 \includegraphics[width=1\columnwidth,angle=0,scale=0.75]{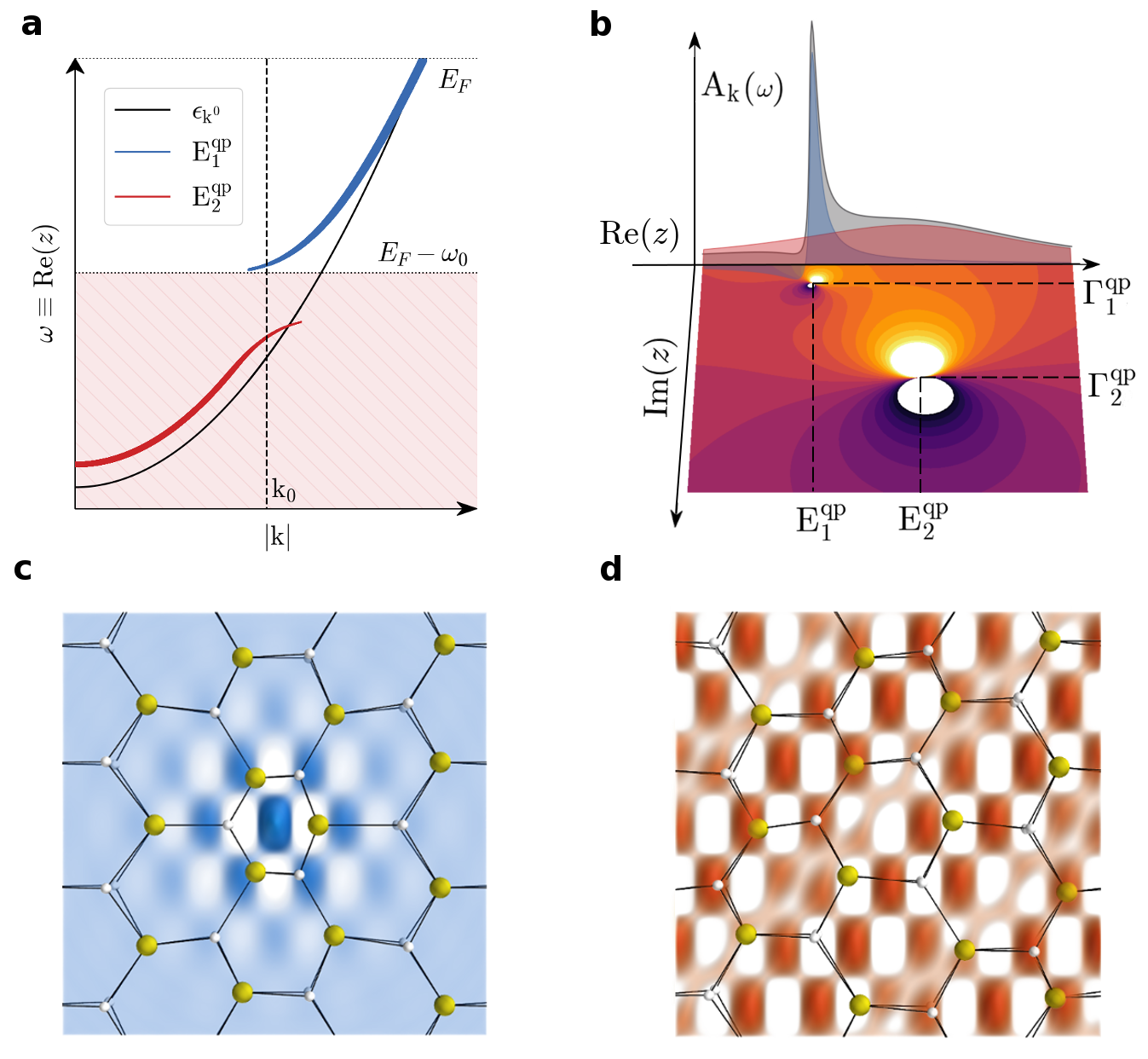}
\end{center}
\caption{ \textbf{Schematic picture of a coupled electron-phonon system.}
\textbf{a,} A bare electron with a parabolic dispersion, represented by a solid 
black line ($\epsilon_{\text{k}}$), interacts with a dispersionless phonon,
represented by a dotted line ($\omega_{0}$).
The electron decay processes 
by phonon emission are energetically allowed only for electrons above $\omega_{0}$ (red background). Therefore, 
the situation is completely different for electrons below this energy range, though even there, quantum field laws allow
for virtual excitations of phonons. The result is that the coupling produces 
two excitation branches $\text{E}^{\text{qp}}_{1}$ and 
$\text{E}^{\text{qp}}_{2}$, whose dispersions are represented by blue and red solid lines, respectively.
Fixing the electron momentum at momentum at $\text{k}_{0}$ (dashed line in \textbf{a}), one obtains a spectral function 
with two peaks at real frequencies (\textbf{b}) corresponding to the above pair of bands. Looking at the complex frequency 
plane, these two excitations are traced back to poles of the electron Green's function 
and their position in the complex plane determines the renormalized energy ($\text{E}^{\text{qp}}$) and lifetime 
broadening ($\Gamma^{\text{qp}}$).
When most of the  spectral function (coherent part) is recovered from 
the superposition of the separate contributions of the poles, 
one may say 
that a multiple quasi-particle picture 
is valid. 
\textbf{c,} 
Electrons below $\omega_{0}$ are strongly renormalized and tend to localize as they are followed
 by a dense virtually emitted phonon cloud. 
\textbf{d,}
On the contrary, the higher energy band is energetically allowed to emit phonons and acquires a more extended character and
a lighter effective mass.} 
\label{fig:Fig1}
 \end{figure*}

The effective velocity and the lifetime of electron states close to the Fermi level determine most of the
transport properties of metals, and the interactions with collective excitactions e.g. 
 phonons, magnons or plasmons, are responsible for modifying or renormalizing these properties~\cite{mahan}. 
Specifically, phonons are  the low
energy excitations that more strongly couple to electron states in normal metals~\cite{grimvall}. 
The interaction of electrons and phonons has a many-body 
character primarily because the Pauli exclusion principle prohibits the scattering to occupied states and because 
quantum mechanics allows the virtual excitation of phonons even in the absence of available energy for low energy electrons.
All this physics is already contained in the most drastically simplified Einstein model, where one single optical phonon mode 
with energy $\omega_0$ interacts 
with a single electron band with a parabolic dispersion in absence of coupling (Fig.\,\ref{fig:Fig1}\textbf{a})\cite{ES}. The many-body 
coupling divides the spectrum in two regions, below and above $\omega_0$. 
On the one hand, for  energies below $\omega_0$, electrons
do not have sufficient energy for emitting any phonon, and therefore they appear long lived. 
However, they are allowed to emit and reabsorb phonons in virtual processes, 
which produces an entire phonon cloud around electrons having the effect of augmenting their effective mass, similar to \textit{polaron} states
in insulators~\cite{devreese_polarons}. 
On the other hand,  for electrons above $\omega_0$ the emission of phonons 
is now allowed, the effect being that
virtual processes are less probable in favour of real emission of phonons leading to a  decreasing of their lifetime and effective mass. 
This idealized picture will  
be useful for understanding the more intricate situation in MoS$_2$
and questions the simple view about quasi-particle properties themselves, since the interaction with phonons 
produces two different electron states with radically different properties.  
A bit more technically, as the energies of quasi-particle states are determined by the poles of the electron propagation amplitude or Green's function 
($G(\mathbf{k},z)$)~\cite{galitskii_migdal},  the presence of several poles in this 
function evidences the existence of as many quasi-particle states (Fig.\,\ref{fig:Fig1}\textbf{b}), 
with very different physical properties in terms of lifetime and dispersion (Fig.\,\ref{fig:Fig1}\textbf{c-d}).
Similarly, in a simple system consisting of two optical phonon modes with different energies, the electron band would split
twice when reaching the energy corresponding to each phonon mode.

%
%
There have been many good examples of ARPES measurements on metal surfaces ~\cite{arpesexpmetal1,arpesexpmetal2,epsurface1}, high-temperature superconductors~\cite{Lanzara2001}, and doped polar insulators~\cite{ARPES_TiO2, controlmb2}, where   strong deviations from the single quasi-particle picture 
have been observed and attributed to the electron-phonon interaction.
However, the recent measurements of Kang~ \textit{et al.}~\cite{Kang2018} show that MoS$_{2}$
is probably the first system where a double splitting in the ARPES spectrum has been observed unambiguously, therefore the importance of understanding the physical 
nature of these broken bands.
These results go also hand in hand with the spin-valley locking protected superconductivity found in this system~\cite{scmos21},
and while some experimental works illustrated the superconducting state
as due to the filling of $\overline{\mathrm{K}}$($\overline{\mathrm{K'}}$) valleys~\cite{scmos22,scmos23,scmos24},
this interpretation clashes with the picture which holds that superconductivity emerges as soon as the Fermi energy ($E_{F}$) crosses the bottom of the
$\overline{\mathrm{Q}}$($\overline{\mathrm{Q'}}$) valleys~\cite{scmos25,phsoftmos2,Piatti2018}. 
In fact, the onset of superconductivity coincides with an outstanding enhancement of the electron-phonon coupling strength~\cite{scmos25}, stemming from phonon-mediated intervalley interactions on the Fermi surface (FS)~\cite{scmos25,phsoftmos2,Piatti2018}, also in agreement with the strong softening of several phonon modes observed in the gate-induced superconducting monolayer MoS$_{2}$~\cite{phsoftmos2}.
We analyse these points and more specifically the case of the observed 
multiple band splitting~\cite{Kang2018} in MoS$_{2}$, where 
the experiments seemed to point out a direct interpretation in terms 
of multiple-phonon excitations. 

\begin{figure*}[ht!]
\begin{center}
 \includegraphics[width=1\columnwidth,angle=0,scale=1.0]{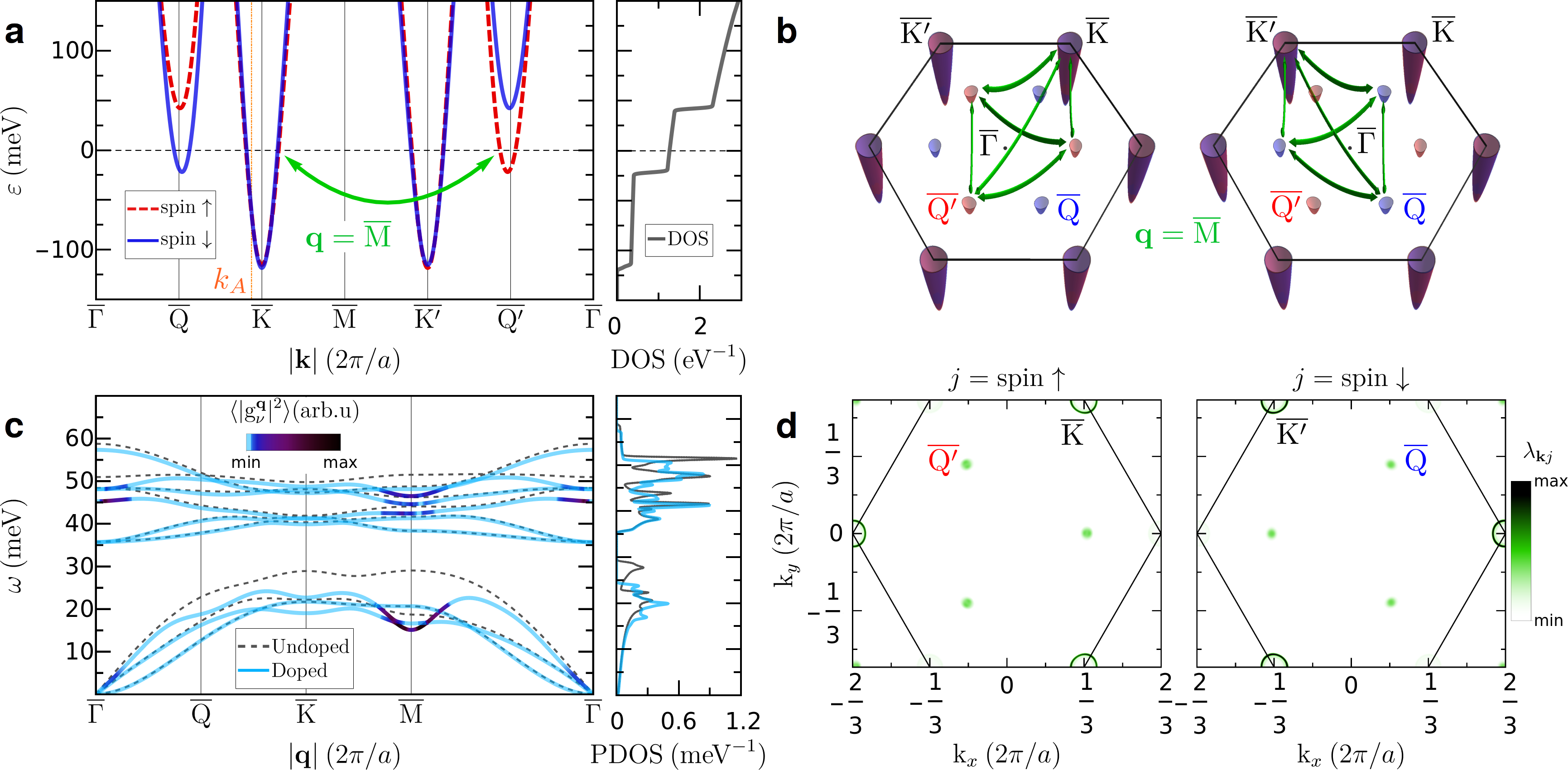}
\end{center}
\caption{   \textbf{Electron-phonon coupling in monolayer MoS$\mathbf{_{2}}$ from first-principles.}
\textbf{a}, Electron band structure (left) and corresponding DOS (right).
Dashed red (spin-up) and solid blue (spin-down) lines represent opposite out-of-plane spin-polarized bands.
$E_{F}$ is set to zero and marked by a horizontal dashed (black) line.
$E_{\overline{\mathrm{K}}(\overline{\mathrm{K'}})}=-118$~meV is the binding energy of the 
conduction-band minima at $\overline{\mathrm{K}}$($\overline{\mathrm{K'}}$).
$E_{\overline{\mathrm{Q}}(\overline{\mathrm{Q'}})}=-22$~meV is the binding energy of the 
minima at $\overline{\mathrm{Q}}$($\overline{\mathrm{Q'}}$).
While $\overline{\mathrm{K}}$($\overline{\mathrm{K'}}$) valleys are almost degenerated, SO interaction induces an energy splitting of $\Delta_{\mathrm{SO}}=70$~meV at
$\overline{\mathrm{Q}}$($\overline{\mathrm{Q'}}$) band edges.
$k_{A}$ momentum is represented by a vertical dotdashed (orange) line.
\textbf{b}, FS of doped $\mathrm{MoS_{2}}$ in the Brillouin zone (BZ) including a 3D-representation of the conduction-band electron pockets centered at the
$\overline{\mathrm{K}}$($\overline{\mathrm{K'}}$) and $\overline{\mathrm{Q}}$($\overline{\mathrm{Q'}}$) points.
Green arrows depict in the left (right) panel the relevant electronic spin-conserving intervalley transitions
driven by phonons with equivalent momenta close to $\mathbf{q}=\overline{\mathrm{M}}$,
connecting $\overline{\mathrm{Q'}}$($\overline{\mathrm{Q}}$) pockets with themselves and with
outer states in $\overline{\mathrm{K}}$($\overline{\mathrm{K'}}$) valley 
for spin-up (spin-down) polarization.
\textbf{c}, Phonon dispersion relation (left) and corresponding phonon DOS (PDOS) (right).
The undoped and doped phonons are represented by dashed (gray) and solid (color code) lines, respectively.
The color code depicts the electron-phonon weighted nesting function ($\langle|g_{\nu}^{\mathrm{q}}|^{2}\rangle_{\mathrm{FS}}$)
for each phonon state $\ket{\mathbf{q},\nu}$ (see Methods).
\textbf{d}, BZ-resolved mass enhancement parameter $\lambda_{\mathbf{k}j}$ of the occupied conduction-states for the spin-up band (left)
and the spin-down band (right). The color code depicts the value of the strength.}
\label{fig:Fig2}
 \end{figure*}

%
%
We made an in-depth theoretical analysis of
the electron-phonon interaction in electron-doped monolayer MoS$_{2}$ by means of first-principles calculations 
in order to shed light on several aspects of the electron-phonon coupling in this system (see also Methods).
We chose a doping carrier density of $n_{\mathrm{2D}}=9\times10^{13}~\mathrm{cm^{-2}}$, 
for which the conduction-band minima at $\overline{\mathrm{K}}$($\overline{\mathrm{K'}}$) are filled and almost 
spin-degenerate (Fig.\,\ref{fig:Fig2}\textbf{a}) with a binding energy of
 $E_{\overline{\mathrm{K}}(\overline{\mathrm{K'}})}=-118$~meV,
 while only the lower spin-split states are populated at $\overline{\mathrm{Q}}$($\overline{\mathrm{Q'}}$) with a binding energy of $E_{\overline{\mathrm{Q}}(\overline{\mathrm{Q'}})}=-22$~meV, which is within the phonon energy range.
The DOS increases step-like as the 2D quasi-parabolic conduction-bands get populated,
showing a noticeable enhancement with the filling of the $\overline{\mathrm{Q}}$($\overline{\mathrm{Q'}}$) bands (Fig.\,\ref{fig:Fig2}\textbf{a}),
the main consequence being that, among all the possible intervalley scattering channels connecting the Fermi sheets (see Supplementary Information), 
the phonon modes with momentum $\mathbf{q}=\overline{\mathrm{M}}$ are the ones dominating
 the whole electron-phonon coupling 
(Fig.\,\ref{fig:Fig2}\textbf{b}). 
This is supported by the large value of the spin-conserving electron-phonon matrix elements (Fig.\,\ref{fig:Fig2}\textbf{c})
for the in-plane polarized acoustic (A) and optical (O) modes with frequencies around $\omega_{\mathrm{A}}^{\overline{\mathrm{M}}}=16$~meV
and $\omega_{\mathrm{O}}^{\overline{\mathrm{M}}}=46$~meV, respectively, which show the largest softenings
comparing to the undoped vibrational spectrum~\cite{phsoftmos2}.
Furthermore, the momentum-resolved mass enhancement parameter $\lambda_{\mathbf{k}j}$ (Methods) for the two occupied spin bands (Fig.\,\ref{fig:Fig2}\textbf{d})
definitely confirms that the electron-phonon coupling is governed by phonons with $\mathbf{q}=\overline{\mathrm{M}}$, yelding  $\lambda_{\mathbf{k}j}$  values as large as 1.2 near $\overline{\mathrm{K}}(\overline{\mathrm{K'}})$.
%
%
%

\begin{figure*}[ht!]
\begin{center}
 \includegraphics[width=1\columnwidth,angle=0,scale=1]{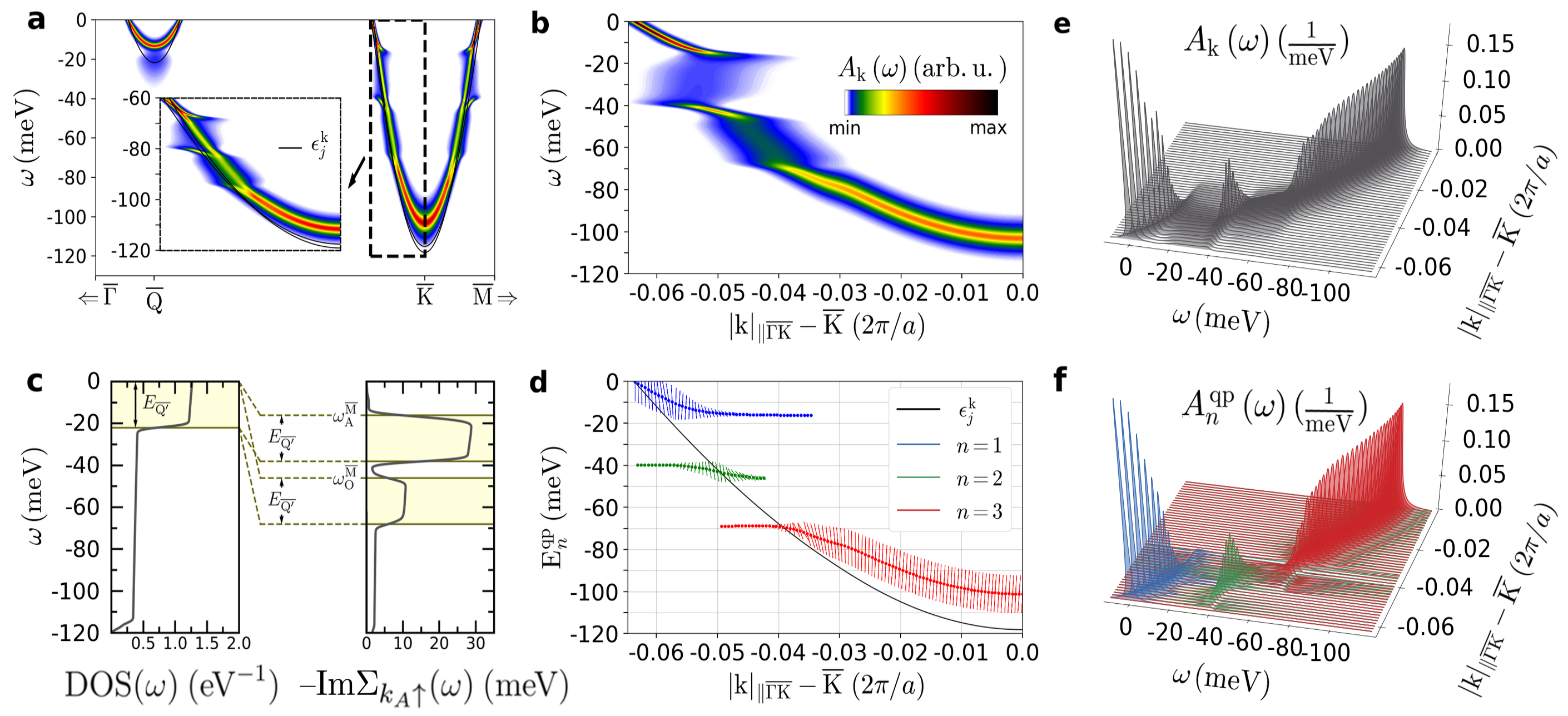}
\end{center}
\caption{  \textbf{Multiple quasi-particle spectra on monolayer MoS$\mathbf{_{2}}$.}
  \textbf{a,} Spectral function of monolayer MoS$_{2}$, calculated from first-principles including electron-phonon interaction effects. The solid black lines represent the non-interacting electron bands.
The dashed rectangle highlights the area of the BZ where the strongest renormalization of the electronic bands occur.
  \textbf{b,} Zoom of the spectral function on the area highlighted in \textbf{a}
 with  the same color code. 
%
\textbf{c,} Imaginary part of the electron self-energy ($\text{Im}\Sigma(\omega)$) (right panel) for an electron with spin-up and momentum $k_{A}$ close to $\overline{\mathrm{K}}$ (Fig.\,\ref{fig:Fig2}\textbf{a}).
The onsets of the rectangular maxima are at $\omega^{\overline{\mathrm{M}}}_{\mathrm{A}}$ and $\omega^{\overline{\mathrm{M}}}_{\mathrm{O}}$,
the energies of the acoustic and optical phonons at $\mathbf{q}=\overline{\mathrm{M}}$, while their width is related to the enhanced DOS at the occupied $\overline{\mathrm{Q'}}$ pockets (yellow shaded area in left panel).
  \textbf{d,} Dispersion of the three quasi-particle poles found for the outer band.
The real part of the poles -- quasi-particle energies -- with respect to the momentum are shown by the blue ($n=1$), green ($n=2$) and red ($n=3$) dots, respectively.
The length of the bars represent the spectral weight of each pole, given by the real part of their residues $\mathbb{Z}_{n}^{\text{qp}}$,
while the imaginary part of the residues are represented by the rotation of the bars, $\text{Im}(\mathbb{Z}_{n}^{\text{qp}})=1$
giving a rotation of $\theta_{n}^{\text{qp}}=\pi$ radians.
  \textbf{e,f} Comparison between the full \textit{ab initio} spectral function (see Eq.\,\ref{eq:specf} in Methods) (\textbf{e}),
 and the contribution coming from each complex quasi-particle pole (\textbf{f}), shown by different colors following the same convention as in \textbf{d}.}
\label{fig:Fig3}
 \end{figure*}

 %
 We have computed the \textit{ab initio} electron spectral function $A(\mathbf{k},\omega)$ including the 
electron-phonon effects without any adjustable parameter (Methods). This function 
displays two sharp band-splittings for electron momenta close to $\overline{\mathrm{K}}$
 at $\omega_{\mathrm{A}}^{\overline{\mathrm{M}}}$ and $\omega_{\mathrm{O}}^{\overline{\mathrm{M}}}$ binding energies
 and width $22$~meV (Fig.\,\ref{fig:Fig3}\textbf{a}-\textbf{b}), each of these band-splittings evoking  
 the one sketched in Fig.\,\ref{fig:Fig1}\textbf{a} for the simplified Einstein model.
 The calculated spectral function is in
 close agreement with the measurements by Kang \textit{et al.}~\cite{Kang2018} and 
 describes the three spectral bands observed in experiment.
 We find that from the two concentric bands around the $\overline{\mathrm{K}}$ valley, 
 it is the outer one (spin-up)  which shows
 the most intense signatures of the electron-phonon coupling, as easily recognized when considering spin conservation arguments (Fig.\,\ref{fig:Fig2}
 \textbf{d}). This is also the reason why the inner spin-locked state shows
 much weaker spectral features. 
 Let us now focus on the strongly interacting outer spin-split band.
 %
 %
 %
 For this state, the frequency dependent imaginary part of the electron self-energy ($\text{Im}\Sigma_{\mathbf{k}j}(\omega)$)
 at momentum $\mathbf{k}=k_{A}$ close to $\overline{\mathrm{K}}$ (Fig.\,\ref{fig:Fig2}\textbf{a})
 shows a rather uncommon rectangular shaped double structure of width $\sim|E_{\overline{\mathrm{Q'}}}|=22$~meV separated by
 a narrow window at $\omega\sim 42$~meV with almost vanishing value (Fig.\,\ref{fig:Fig3}\textbf{c} right).
 These two rectangular shapes have onsets precisely at $\omega_{\mathrm{A}}^{\overline{\mathrm{M}}}$ and $\omega_{\mathrm{O}}^{\overline{\mathrm{M}}}$ 
 binding energies and they are easily rationalized in terms of energy conservation for the corresponding phonon emission processes
 connecting states close to $\overline{\mathrm{Q'}}$ with those near $\overline{\mathrm{K}}$ (Fig.\,\ref{fig:Fig2}\textbf{b}).
 It is actually the large DOS at the occupied $\overline{\mathrm{Q'}}$ pockets (Fig.\,\ref{fig:Fig3}\textbf{c} left)
 which enhances the phase-space of these strongly-interacting scattering processes,
 yielding the maximum values of $\text{Im}\Sigma_{k_{A}\uparrow}(\omega)$ (yellow shaded areas in Fig.\,\ref{fig:Fig3}\textbf{c}).
 The dip in the imaginary part of the self-energy appears as long as the condition
 $\omega_{\mathrm{O}}^{\overline{\mathrm{M}}}-\omega_{\mathrm{A}}^{\overline{\mathrm{M}}}>|E_{\overline{\mathrm{Q'}}}|$ holds,
 since the quasi-particles near $\mathbf{k}=k_{A}$ with energies between the two maxima can not scatter to $\overline{\mathrm{Q'}}$ valleys.
 %
 %
 This means that the most uncommon spectral feature found in ARPES close to $\sim 42$~meV, 
also found in  our calculations (Fig.\,\ref{fig:Fig3}\textbf{b}), 
 if demonstrated to be a  well-defined quasi-particle, it would correspond to a  very long-lived state  
even if it is strongly  interacting. This is a unique characteristic of MoS$_2$, arising from the fact that  
 the relevant doped valleys have different binding energies and that there are basically two
 relevant phonon modes connecting precisely 
 those valleys. 
Recall that in ordinary metals, where the DOS is practically constant for typical phonon energies, 
 the imaginary part of the electron-phonon self-energy is a monotonically increasing function, which 
is radically different to what is observed in MoS$_2$. Note that thermal and electron-electron interactions are
expected to slightly soften the spectral features obtained considering only the electron-phonon coupling.
An estimation of the electron-electron 
interaction considering RPA~\cite{RPA} adds a slowly monotonically increasing function 
yielding a broadening of $\Gamma_{ee} \sim 2.8 ~\text{meV}$ even for electron energies close to $\omega \sim 40$~meV. 

%
%


It's worth looking a bit more closely at the general properties of the quasi-particles even if it's succinctly, in order 
to see if the observed spectral features can be defined as actual quasi-particles. These
find their appropriate mathematical definition as poles of the electron Green's function 
$G_{\mathbf{k}}(z)=1/\left(z - \epsilon_{j}^{\mathbf{k}} - \Sigma_{\mathbf{k}j}(z)\right )$, that is, the solutions of 
the so-called 
Dyson equation $z - \epsilon_{j}^{\mathbf{k}} - \Sigma_{\mathbf{k}j}(z) = 0$. The key point here is that 
due to the nonlinear character of this equation it may lead to several solutions, even for a single unperturbed
 state with energy  
$\epsilon_{j}^{\mathbf{k}}$. The consistent use of the complex plane allows to treat the Green's
 function in the entire plane
and the energy ($\mathrm{E^{\text{qp}}}$) and the lifetime broadening ($\Gamma^{\text{qp}}$) of possible quasi-particles 
are compactly 
expressed as an ordinary complex number $z_{n}^{\text{qp}} = \mathrm{E^{\text{qp}}} - i \Gamma^{\text{qp}}$.   
Then the Dyson equation leads to a pair of coupled non-linear equations~\cite{asierprb2009},
\begin{equation}\label{eq:qpeq}
 z_{n}^{\text{qp}} - \epsilon_{j}^{\mathbf{k}} - \Sigma_{\mathbf{k}j}( z_{n}^{\text{qp}}) = 0 \rightarrow 
\left \{ \begin{array}{c}
 \mathrm{E}^{\text{qp}}_{n}-\epsilon_{j}^{\mathbf{k}}-\text{Re}\left ( \Sigma_{\mathbf{k}j}(\mathrm{E}^{\text{qp}}_{n} - i \Gamma^{\text{qp}}_{n} ) \right ) = 0 \\
         \Gamma^{\text{qp}}_{n} + \text{Im}\left ( \Sigma_{\mathbf{k}j}(\mathrm{E}^{\text{qp}}_{n} - i \Gamma^{\text{qp}}_{n} ) \right ) = 0
\end{array} \right \}.
\end{equation}
Capturing the influence of the quasi-particle lifetime broadening on its shifted energy
and vice versa, as above, is not the most standard procedure but appears absolutely crucial 
when trying to understand the spectral features of many strongly interacting systems 
in terms of elementary excitations, as shall be the case also in MoS$_2$.

%
%
We have solved Eq.(\ref{eq:qpeq}) for monolayer MoS$_{2}$, and the results are shown in Fig.\ref{fig:Fig3}\textbf{d}. 
For the sake of clarity, we focused on the same momentum region as in Fig.\ref{fig:Fig3}\textbf{b}, 
and only on the strongly interacting outer spin-split band.
Close to the Fermi momentum $\textit{k}_{\textit{F}}$, an Engelsberg-Schrieffer-like~\cite{ES}  state appears with a strongly renormalized 
dispersion, and it is denoted as the $n=1$ solution. Far enough from $\textit{k}_{\textit{F}}$, we find a dispersive and damped state  
identified as the $n=3$ solution.
An important conclusion is that for some intermediate values of the momentum we find an additional solution ($n=2$)
 with an important 
spectral weight which is practically flat, and it is therefore a strongly interacting state
which tends to localization. 
However, this state appears long lived 
as it lies just in the energy window where the imaginary part of the self-energy has almost a gap 
(Fig.\ref{fig:Fig3}\textbf{c}). More specifically, the electron-phonon limited lifetime broadening at this energy range
is almost negligible $\Gamma^{\text{qp}}_{n=2} \sim 0.35~\text{meV}$.
%
Considering the Dyson equation in the complex plane allows also to associate a precise spectral weight
 to each quasi-particle state, in a  systematic way, 
and is simply given by the residue of the poles ($\mathbb{Z}_{n}^{\text{qp}}=1/\left ( 1-\Sigma^{'}(z_{n}^{\text{qp}}) \right )$) evaluated at 
complex quasi-particle energies. It is not therefore necessary to visually analyse the spectral function $A(\omega)$ in order to check 
for the energies 
of possible quasi-particle states and/or their relative importance, 
as these are given directly by $z^{\text{qp}}$ and $\mathbb{Z}_{n}^{\text{qp}}$. 
This allows to  define the coherent part of the spectral-function systematically
($A^{\text{qp}}(\mathbf{k},\omega) = - \sum_n \frac{1}{\pi} ~ 
\text{Im}\left ( \frac{\mathbb{Z}_{n}^{\text{qp}}(\mathbf{k})}{\omega - z_{n}^{\text{qp}}(\mathbf{k})}\right)~$), and explicitly broken down 
into separate contributions from each quasi-particle pole. Note for a moment 
the severe consistency condition $\sum_n \text{Re} \left ( \mathbb{Z}_{n}  \right ) \lesssim 1$ imposed here
as the integral of $A^{\text{qp}}(\mathbf{k},\omega)$ must  be of the order but less than unity. 
Said in pass, this result 
is not obtained by any means when only the real part of the self-energy is considered for calculating the 
dispersion of quasi-particles (see Supplementary Information). 
When the contribution of each separate pole $A_{n=1,3}^{\text{qp}}(\mathbf{k},\omega)$ is plotted 
as in Fig.\ref{fig:Fig3}\textbf{f}, the comparison with the full spectral function (Eq.\ref{eq:specf} in Methods)
 in Fig.\ref{fig:Fig3}\textbf{e} 
is strikingly good. In this figure it is also shown how strongly the spectral weight from one quasi-particle into others  is transferred
as a function of $\textbf{k}$, even when for some values of momentum all the three many-body solutions are present.
Altogether, the obtained quasi-particle band structure -in its complex version- almost perfectly resembles the three-peak and double-gap structure 
observed in ARPES. Therefore, the threefold band structure observed in experiment corresponds
 certainly to quasi-particle states and not to multiple-phonon (high order) processes nor 
 to side-bands or satellites without a clear physical meaning. 

These findings provide a theoretical explanation for the singular spectral features observed on monolayer 
MoS$_{2}$~\cite{Kang2018} in terms of three elementary many-body quasi-particle states.
Close to the $\overline{\mathrm {K}}$ valley, the available scattering phase-space appears severely
restricted for some narrow 
energy windows, with the result that one quasi-particle branch ($n=2$) appears exceptionally long lived, even 
when the strong coupling induces a practically flat dispersion for this band. Flat dispersion indicates a sort
of real space localization property 
of these states and, therefore, the effective interactions between them should be expected to be profoundly modified.
As the transport and superconducting properties~\cite{scmos21,scmos22,scmos23,scmos24} depend so dramatically 
on the lifetime, dispersion and effective interactions between elementary excitations, we believe that these
results may serve as a guidance to understand, explore and eventually take adavantage of many-body interactions.

\begin{methods}

\subsection{First-principles calculations.}
The first-principles calculations were performed within the noncollinear density functional theory (DFT)~\cite{DFT1,DFT2} and density functional perturbation theory (DFPT)~\cite{DFPT}
with fully relativistic norm-conserving pseudopotentials as implemented in the \textsc{Quantum Espresso} package~\cite{QE} 
and using the Perdew-Zunger local density approximation parametrization for the exchange-correlation functional~\cite{PZLDA}.
In order to correctly describe the ground-state electronic and phononic structures of the single-layer MoS$_{2}$,
we used the experimentally measured in-plane bulk lattice parameter $a=3.16$~{\AA}~\cite{aexp} and a vacuum layer of five times the lattice parameter,
which is large enough for avoiding any interplay between adjacent layers.
Carrier doping effects were simulated by the addition of excess electronic charge into the unit cell, which is compensated by a uniform positive jellium background.
All atomic forces were relaxed up to at least $10^{-6}$~ Ry/a.u..
We used a $36\times36$ Monkhorst-Pack grid for the self-consistent electronic calculations, while the lattice vibrational properties were evaluated on a coarse
$9\times9$ $q$-point mesh.

\subsection{Electron-phonon interaction calculations.}
The self-consistent electron-phonon magnitudes were computed considering doping-sensitive full-spinor electron states,
as well as doping-sensitive phonon states and deformation potentials. In a first step we considered a coarse mesh of
$9\times9$ for phonons $(\mathbf{q})$ and a $18\times18$ mesh for electron states $(\mathbf{k})$, but 
for fine integrals, we used the Wannier interpolation scheme~\cite{giustinoreview2017,giustinowannier,asierwannier}, which allowed us to consider $10^7$ points in the Brillouin zone for electrons and $10^6$ for phonons. 
The Fermi surface integrated squared matrix elements (Fig.\,\ref{fig:Fig2}\textbf{c}),
 equivalent to a sort of electron-phonon weighted nesting function defined
for the phonon branch ($\nu$) at $\mathbf{q}$,  is defined by the following integral: 
 \begin{equation}
  \langle|g_{\nu}^{\mathbf{q}}|^{2}\rangle_\mathrm{FS}=\frac{1}{N(E_{F})} \sum_{\mathbf{k}ij} |g_{ij}^{\nu}(\mathbf{k},\mathbf{q})|^2
  \delta(\epsilon_{j}^{\mathbf{k}}-\epsilon_{i}^{\mathbf{k}+\mathbf{q}}\pm\omega_{\nu}^{\mathbf{q}}) \delta(\epsilon_{j}^{\mathbf{k}}-E_{F}),
  \label{eq:gqv2}
 \end{equation}
where $\epsilon_{j}^{\mathbf{k}}$ is the single-particle bare eigenvalue of the electron state with  momentum
$\mathbf{k}$ in the band $j$,
$\omega_{\nu}^{\mathbf{q}}$ is the frequency of the lattice vibrational normal mode with momentum $\mathbf{q}$ and mode $\nu$,
$N(E_{F})$ the DOS at $E_{F}$,
and $g_{ij}^{\nu}(\mathbf{k},\mathbf{q})$  the electron-phonon matrix element connecting 
the $\ket{\mathbf{k},j}$ and $\ket{\mathbf{k}+\mathbf{q},i}$ electron states by a phonon $\ket{\mathbf{q},\nu}$.
This quantity allows us to identify the relevant phonon modes interplaying with electrons at $E_{F}$
and offers a measure of the strength of the coupling for each phonon mode $\ket{\mathbf{q},\nu}$.

The mass enhancement parameter or  Fermi surface averaged electron-phonon coupling strength is defined as~\cite{grimvall}:
 \begin{equation}
  \lambda=\frac{1}{N(E_{F})}\frac{1}{N_{q}} \sum_{\mathbf{k}\mathbf{q}ij\nu} \frac{|g_{ij}^{\nu}(\mathbf{k},\mathbf{q})|^2}{\omega_{\nu}^{\mathbf{q}}}
  \delta(\epsilon_{j}^{\mathbf{k}}-\epsilon_{i}^{\mathbf{k}+\mathbf{q}}\pm\omega_{\nu}^{\mathbf{q}}) \delta(\epsilon_{j}^{\mathbf{k}}-E_{F}).
  \label{eq:lambda}
 \end{equation}
This factor reveals how much the effective mass is enhanced $m=m_0(1+\lambda)$ at the Fermi level due to the electron-phonon 
interaction.
The state $\ket{\mathbf{k},j}$  resolved  electron-phonon coupling strength (Fig.\,\ref{fig:Fig2}\textbf{d}) was evaluated as:
 \begin{equation}
  \lambda_{\mathbf{k}j}=\frac{1}{N_{q}} \sum_{\mathbf{q}i\nu} \frac{|g_{ij}^{\nu}(\mathbf{k},\mathbf{q})|^2}{\omega_{\nu}^{\mathbf{q}}}
  \delta(\epsilon_{j}^{\mathbf{k}}-\epsilon_{i}^{\mathbf{k}+\mathbf{q}}\pm\omega_{\nu}^{\mathbf{q}}).
  \label{eq:lambdakj}
 \end{equation}

The semi-empirical McMillan-Allen-Dynes formula for estimating the superconducting critical temperature $T_{\mathrm{c}}$ is defined as,
 \begin{equation}
  T_{\mathrm{c}}=\frac{\omega_{log}}{1.2}\mathrm{exp}\Bigg(-\frac{1.04(1+\lambda)}{\lambda-\mu^{*}(1+0.62\lambda)}\Bigg),
  \label{eq:tc}
 \end{equation}
where $\omega_{log}$ is a logarithmic average of the phonon frequencies and $\mu^{*}$ is a parameter describing the Coulomb repulsion.
We estimated the superconducting critical temperature of the doped monolayer MoS$_{2}$ for some values
 of $\mu^{*}$ within the range of $0.1-0.3$, obtaining values in the range of $4-8$~ K, with a calculated electron-phonon coupling 
of $\lambda \sim 0.63$.
This estimation is in agreement with the experimentally measured critical temperature of $8.5$~K
at a 2D-carrier density $n_{\mathrm{2D}}=9\times10^{13} \mathrm{cm^{-2}}$~\cite{scmos21,scmos22,scmos23,scmos24}.

\subsection{Electron self-energy and spectral function.}
The \textit{ab inito} expression for the Fan-Migdal electron self-energy valid at $T=0$~K used in our calculations 
was~\cite{giustinoreview2017}:
 \begin{equation}
  \Sigma_{\mathbf{k}j}(\omega)=\frac{1}{N_{q}}\sum_{\mathbf{q}i\nu}|g_{ij}^{\nu}(\mathbf{k},\mathbf{q})|^{2}
  \Bigg( \frac{f(\epsilon_{i}^{\mathbf{k}+\mathbf{q}})}{\omega-\epsilon_{i}^{\mathbf{k}+\mathbf{q}}+\omega_{\nu}^{\mathbf{q}}+i\eta} +
         \frac{1-f(\epsilon_{i}^{\mathbf{k}+\mathbf{q}})}{\omega-\epsilon_{i}^{\mathbf{k}+\mathbf{q}}-\omega_{\nu}^{\mathbf{q}}+i\eta}\Bigg),
  \label{eq:se}
 \end{equation}
where $\omega$ represents the quasi-particle energy in the real-energy axis, $f$ denotes the Fermi-Dirac occupation factor, 
and $\eta$ is a real positive infinitesimal. 
%
%
The electron self-energy as calculated in Eq.\,\ref{eq:se}  allows to obtain the spectral function  
for real frequencies~\cite{mahan,grimvall,giustinoreview2017}:
 \begin{equation}
 \begin{aligned}
  A(\mathbf{k},\omega)&=-\frac{1}{\pi}\sum_{j}\text{Im}\left ( G(\mathbf{k},\omega) \right ) 
                      =-\frac{1}{\pi}\sum_{j}\text{Im}\left ( \frac{1}{\big(\omega-\epsilon_{j}^{\mathbf{k}}-\Sigma_{\mathbf{k}j}(\omega)\big)}
                      \right ) \\
                      &=-\frac{1}{\pi}\sum_{j}\frac{\text{Im}\left( \Sigma_{\mathbf{k}j}(\omega) \right )}{\left(\omega-\epsilon_{j}^{\mathbf{k}}-\text{Re}\left ( \Sigma_{\mathbf{k}j}(\omega) \right  )\right)^2+\left (\text{Im} \left ( \Sigma_{\mathbf{k}j}(\omega) \right )\right)^2}.
    \label{eq:specf}
 \end{aligned}
 \end{equation}

\subsection{Analytic continuation of the electron self-energy.}

In order to find solutions of the complex quasi-particle equation Eq.\,\ref{eq:qpeq}, we need to perform an analytical continuation of the electron self-energy $\Sigma(\omega)$ from the upper half complex plane into the lower half.
First, we calculated the self-energy from first principles only for real frequencies (Eq.\,\ref{eq:se}), and next
we generalized the method outlined in Ref.\,\cite{asierprb2009} to handle self-energies without assuming particle-hole symmetry.
For doing so, we used the Kramers-Kronig relation integrated by parts,
\begin{equation}\label{eq:Kramers-Kronig}
 \Sigma(\omega + i\delta) = \int_{-\infty}^{\infty} d\omega' ~ \frac{d~\mathrm{Im}\left(\Sigma(\omega'')\right)}{d~\omega''}\Bigr|_{\omega''=\omega'} ~ \log (\omega - \omega' + i\delta)~.
\end{equation}
As it is known, a direct substitution of $\omega$ by $z$ in Eq.\,\ref{eq:Kramers-Kronig} is not valid, as the branch-cuts introduced by the $\log$ term makes the direct numerical integration inappropriate. However, a piecewise polynomial interpolation of 
$\frac{d\mathrm{Im}\left(\Sigma(\omega)\right)}{d\omega}$ and a subsequent analytical integration is a possible numerical strategy. 
We used a cubic spline interpolation, ensuring the continuity of the first and second derivatives all over the $\omega$ axis.

\end{methods}




\bibliographystyle{naturemag}
\bibliography{article}


\begin{addendum}
\item 
The authors acknowledge the
Department of Education, Universities and Research of the
Basque Government and UPV/EHU (Grant No. IT756-13)
and the Spanish Ministry of Economy and Competitiveness MINECO 
(Grant No.  FIS2016-75862-P) for financial support. 
 P.G. and J.L. acknowledge financial support from the
University of the Basque Country UPV/EHU (Grant Nos.
PIF/UPV/12/279 and  PIF/UPV/16/240, respectively) and
the Donostia International Physics Center (DIPC)
for financial support.
Computer facilities
were provided by the DIPC.
\item[Competing Interests] The authors declare that they have no
competing financial interests.
\item[Correspondence] Correspondence and requests for materials
should be addressed to A.E.~(email: asier.eiguren@ehu.eus).
\end{addendum}


\clearpage

\end{document}